\documentclass[lettersize,journal]{IEEEtran}
\usepackage{amsmath,amsfonts}
\usepackage{algorithmic}
\usepackage{algorithm}
\usepackage{array}
\usepackage{textcomp}
\usepackage{url}
\usepackage{verbatim}
\usepackage{graphicx}
\usepackage{cite}
\usepackage{enumitem}
\usepackage{graphicx}
\usepackage{booktabs}
\usepackage{lipsum}
\usepackage{stfloats}
\usepackage{hyperref}
\usepackage{xcolor}
\usepackage{bm}
\usepackage{amsmath}
\usepackage{booktabs}
\usepackage{multirow}
\usepackage{float}
\usepackage{xurl}
\hypersetup{breaklinks=true}

\usepackage{caption}
\usepackage{subcaption}
\usepackage[labelformat=simple]{subcaption}

\begin{document}
\title{From Multilayer Perceptron to GPT:  A Reflection on Deep Learning Research for \\Wireless Physical Layer}

\author{Mohamed Akrout, Amine Mezghani,~\IEEEmembership{Member,~IEEE}, Ekram Hossain,~\IEEEmembership{Fellow,~IEEE}, \\ Faouzi Bellili,~\IEEEmembership{Member,~IEEE}, Robert W. Heath,~\IEEEmembership{Fellow,~IEEE}
}



\maketitle

\begin{abstract}
Most research studies on deep learning (DL) applied to the physical layer of wireless communication do not put forward the critical role of the accuracy-generalization trade-off in developing and evaluating \textit{practical} algorithms. To highlight the disadvantage of this common practice, we revisit a data decoding example from one of the first papers introducing DL-based end-to-end wireless communication systems to the research community and promoting the use of artificial intelligence (AI)/DL for the wireless physical layer. We then put forward two key trade-offs in designing DL models for communication, namely, accuracy versus generalization and compression versus latency. We discuss their relevance in the context of wireless communications use cases using emerging DL models including large language models (LLMs). Finally, we summarize our proposed evaluation guidelines to enhance the research impact of DL on wireless communications. These guidelines are an attempt to reconcile the empirical nature of DL research with the rigorous requirement metrics of wireless communications systems.

\end{abstract}

\section{Introduction}

\begin{figure}[!h]
\centering
\includegraphics[scale=0.5]{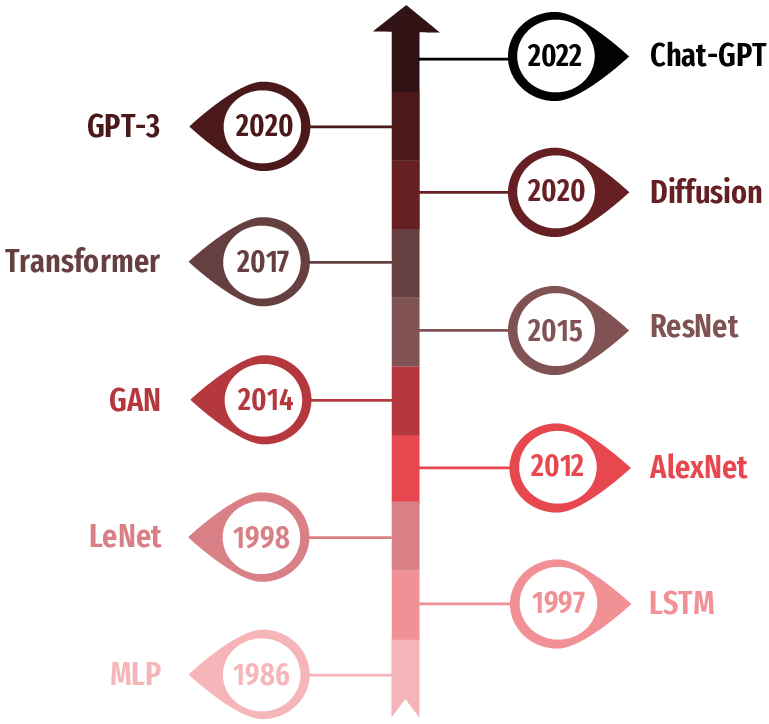}
\caption{Publication timeline of the key DL models used and/or being investigated by the communications research  community.}
\label{fig:timeline-models}
\vspace{-0.2cm}
\end{figure}

\IEEEPARstart{R}{esearchers} are developing use cases where DL can potentially enhance the system performance and reduce complexity/overhead compared to classical methods (see envisioned examples for standardization in the 3GPP Release 18 \cite[Section 9.2]{3GPPRelease18}). Since the introduction of the AlexNet network in 2012, the use of deep neural networks (DNNs) has skyrocketed within the communications community by substituting conventional optimization solvers with generative and/or discriminative DL techniques. Fig. \ref{fig:timeline-models} summarizes some of the key DL models applied to communications problems, starting from the multilayer perception (MLP) model \cite{rumelhart1986learning} to the diffusion model \cite{ho2020denoising}. We refer the reader to \cite{tanveer2022machine} for a comprehensive survey on the applications of DL for wireless physical layer design. In Fig. \ref{fig:timeline-models}, we also include the recent generative pre-trained transformer (GPT) models \cite{brown2020language,openai2023gpt4} as they are currently initiating many  discussions within the communications research community about the data compression properties of large language models (LLMs) and their role in replicating digital twins (cf. Section \ref{subsec:compression-latency}).

Because the sixth-generation (6G) networks are envisaged as multi-band, decentralized, fully autonomous, and hyper-flexible user-centric systems encompassing satellite, aerial, terrestrial, underwater, and underground communications, DL techniques are expected to partially or fully substitute classical methods, their assessment metrics should comply with rigorous evaluation guidelines that equally address latency, complexity, generalization, and accuracy. In this paper, we reflect on the last  decade of research on DL for wireless communications with a focus on the physical and link layers. We start by highlighting the limitations of state-of-the-art DL methods for wireless communications. We do so by revisiting one of the first published papers \cite{o2017introduction} introducing DL-based end-to-end communications systems over additive white Gaussian noise (AWGN) channels. By doing so, we pinpoint how \cite{o2017introduction}, as many of the subsequent DL papers for wireless, turn the spotlight on the accuracy of DL methods at the cost of sacrificing their generalization capabilities. We also highlight how the open literature does not draw enough attention to the important practical considerations for designing DL systems for wireless communications, such as data acquirement and adaptation to new system dimensions. We then describe the two key trade-offs in designing DL-aided wireless communications systems, namely, \textit{accuracy versus generalization} and \textit{compression versus latency}. These two trade-offs offer new evaluation guidelines to assess future DL research directions for wireless communications.\vspace{-0.1cm}

\section{Limitations of the State-of-the-Art\\ DL Techniques for Wireless Communications}

In this section, we describe the limitations of neglecting the data distribution shift when evaluating DL models for wireless communications at the physical layer. Specifically, we revisit an example from \cite{o2017introduction} to illustrate the drawbacks of blindly applying black-box DL models in a plug-and-play manner when only the model accuracy is assessed. We also describe other challenges arising from the use of DL techniques that are usually not examined in the open literature.

\subsection{Evaluation of a Single Metric}\label{subsec:single-metric-eval}

It is always possible to beat a classical method that solves a non-closed form problem using DL techniques based on deep neural networks (DNNs) given a known model by generating training and test datasets on which DNNs are \textit{both} trained and evaluated. For this reason, it is critical to use both the accuracy and generalization of DNNs to assess their performance on a variety of:
\begin{itemize}[leftmargin=*]
    \item in-distribution (ID) scenarios where the training and testing datasets are generated from the same distribution (e.g., the same user speed is assumed to generate mobility data  to be used for training and testing).
    \item out-of-distribution (OOD) scenarios where a distribution shift occurs between training and testing datasets (e.g., different user speeds are assumed to generate mobility data  to be used for training and testing).  
\end{itemize}

\noindent In other words, DL models that are either accurate and non-robust or highly biased and robust are equally worthless for real-time physical/link layer applications.

\begin{figure}[!h]
\centering
\includegraphics[scale=0.37]{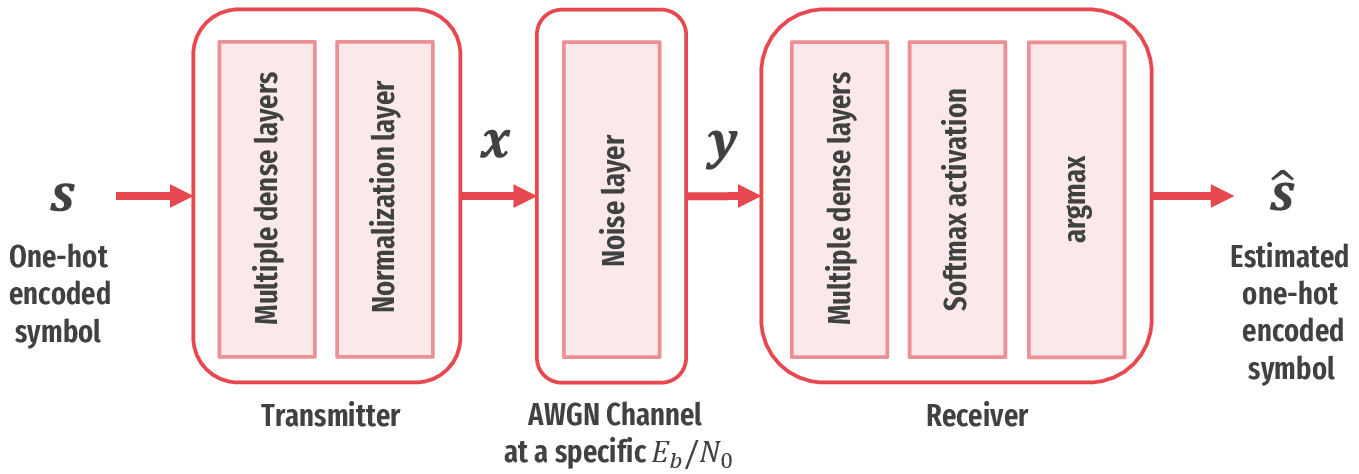}
\caption{An end-to-end communication system over an AWGN channel as an autoencoder: the input is the one-hot encoded transmit symbol $\bm{s}$ while the estimated one-hot symbol $\widehat{\bm{s}}$ is argmax of the softmax probability distribution over all possible messages.}
\label{fig:end2end}
\end{figure}

To illustrate this idea, we revisit the study \cite{o2017introduction}, which is one of the first papers promising the design of communications systems as an autoencoder for reconstruction tasks that jointly optimize transmitter and receiver components in a single end-to-end process. As shown in Fig. \ref{fig:end2end}, the physical communication chain of the transmitter and the receiver are substituted by an encoder and a decoder, respectively. The channel is represented as a noisy non-parameterized layer that injects an additive Gaussian white noise at a specific energy per bit to noise power spectral density ratio, $E_b/N_0$.

\noindent In \cite{o2017introduction}, the training of the autoencoder was performed on a dataset generated at $E_b/N_0 = 7\,\textrm{dB}$. However, the evaluation was conducted over a range of $E_b/N_0$ between in $[-4\,\textrm{dB},\,8\,\textrm{dB}]$. By doing so, the authors obtained a lower block error rate (BLER) than the Hamming code with rate $R=4/7$ and concluded that the autoencoder ``has learned some joint coding and modulation scheme, such that a coding gain is achieved''.

\begin{figure}[h!]
\vspace{-0.2cm}
\centering
\includegraphics[scale=0.27]{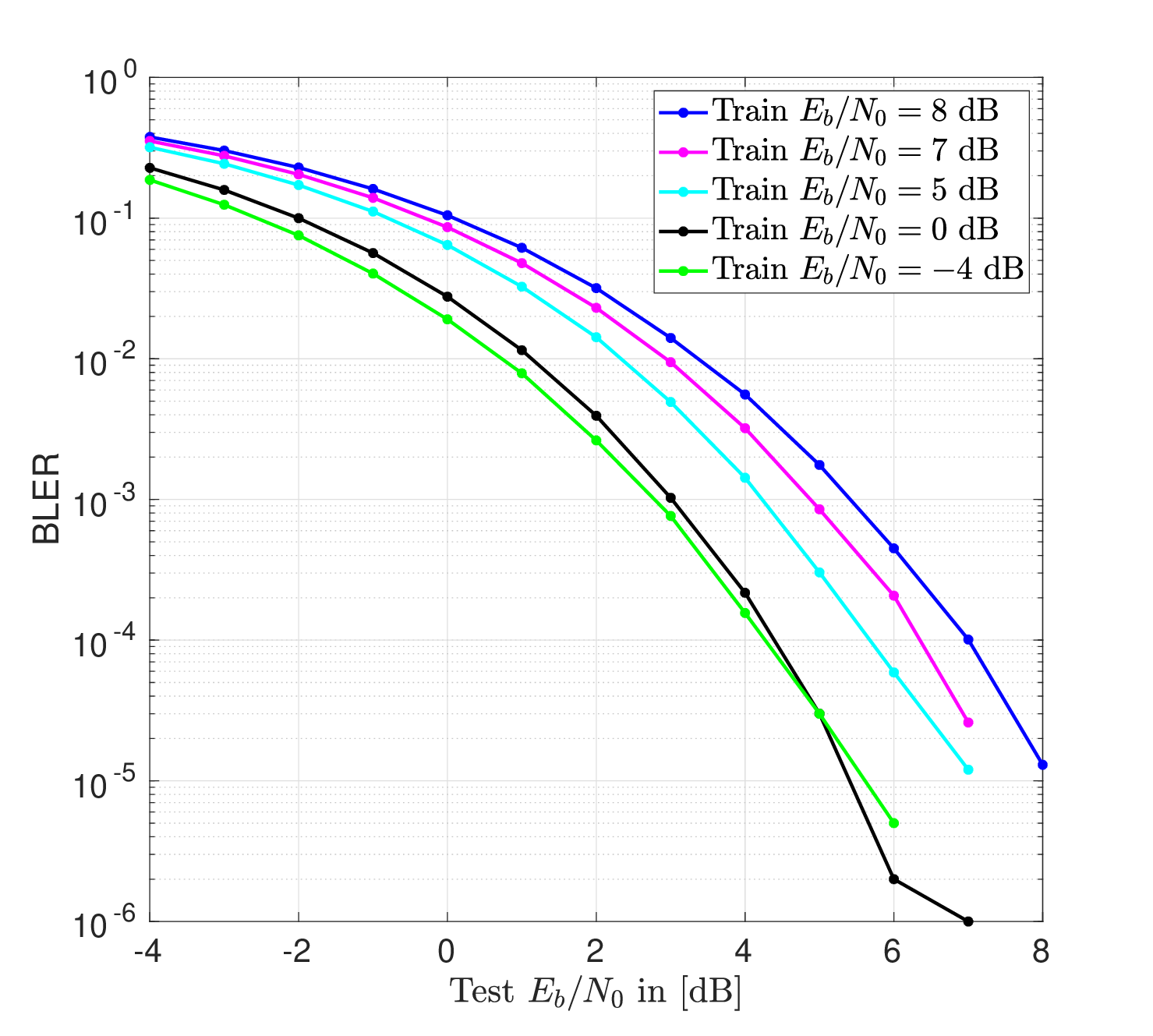}
\caption{BLER of the autoencoder versus test $E_b/N_0$ for multiple training $E_b/N_0$.}
\label{fig:ae}
\end{figure}

From a machine learning theory perspective, this conclusion is questionable. For this reason, we train multiple autoencoders, each with a training $E_b/N_0 \in \{-4,0,5,7,8\}$ dB. We then evaluate each autoencoder on the testing range $E_b/N_0 \in [-4\,\textrm{dB},\,8\,\textrm{dB}]$ as shown in Fig. \ref{fig:ae}. There, all autoencoders in Fig. \ref{fig:ae} exhibit a decreasing BLER over the entire test $E_b/N_0$ range even though they were trained on one single $E_b/N_0$ value. While this fact suggests that autoencoders over AWGN channels generalize well to out-of-distribution decoding scenarios, it can also indicate that the problem at hand is easy to solve because the distribution shift between the training data distribution associated with $E_b/N_0 = 7\,\textrm{dB}$ and the test data distributions with $E_b/N_0 \in [-4\,\textrm{dB},\,8\,\textrm{dB}]$ is minimal. To confirm this fact, we report in Table \ref{table:area-overlap} the percentage of the area overlap  between the data distribution of the received signal $\bm{y}\sim \mathcal{N}\left(\bm{x},\frac{1}{2\,R\,E_b/N_0}\,\mathbf{I}\right)$ for training and testing  $E_b/N_0$ values \footnote{To quantify the shift between two distributions, their area overlap is more informative than their KL-divergence because the latter only accounts for the region where both distributions are non-zero.}. We observe the fact that  lower testing values for $E_b/N_0$ compared to the training value  $E_b/N_0 = 7\,\textrm{dB}$ yields a lower area overlap, or equivalently, a higher OOD shift. However, it is interesting to note the significantly high overall overlap. This suggests that the AWGN channel is a simplistic model to assess the generalization performance of DL models. One can perceive the AWGN channel model in communication as the MNIST dataset in computer vision on which any great performance is considered obsolete by the computer science community due to the intrinsic simplicity of the handwritten digit classification task.
 
\begin{table}[h!]
\caption{The percentage of area overlap between the training ($E_b/N_0 = 7\,\textrm{dB}$) and testing received signal distributions ($E_b/N_0 \in \{-4,0,5,8\}$ dB).}
\centering
\begin{tabular}{ccccc} \toprule
    {\textbf{Test} $\mathbf{E_b/N_0}$} & {$-4$ dB} & {$0$ dB} & {$5$ dB} & {$8$ dB}  \\ \midrule
    \textbf{Area overlap with} & \multirow{2}{*}{45.70\%}&\multirow{2}{*}{62.97\%} & \multirow{2}{*}{88.91\%} & \multirow{2}{*}{94.43\%}\\
    \textbf{train} $\mathbf{E_b/N_0 = 7}$ \textbf{dB} & & & \\ \bottomrule
\end{tabular}
\label{table:area-overlap}
\end{table}

\noindent It is also unclear whether the chosen value $E_b/N_0 = 7\,\textrm{dB}$ is the best one to select. It is seen that training autoencoders on lower $E_b/N_0$ values in $[-4\,\textrm{dB},\,8\,\textrm{dB}]$ (i.e., with higher noise levels) leads to a smaller BLER across the entire test $E_b/N_0$ interval. This result not only shows that the train $E_b/N_0 = 7\,\textrm{dB}$ selected in \cite{o2017introduction} is not the best choice, but also demonstrates how noisy training can be more beneficial for better generalization.

Given all the aforementioned reasons, the fact that the autoencoder outperformed the Hamming code in \cite{o2017introduction} is more justified by the simplicity of the AWGN channel model which does not shift significantly the received signal in the testing $E_b/N_0$ interval. As a matter of fact, adding a noise correlation to the AWGN channel or accounting for the fading effects by changing the AWGN channel to a Rayleigh one break down the decoding performance of the autoencoder. In summary, the analysis of the results as a function of the trade-off between generalization and accuracy metrics opens the door to future rigorous investigations and reveals more insights about better data generation and model training choices.

\subsection{Is Meta-learning Sufficient for Generalization?}
To sidestep the need for large data samples to train DL models, meta-learning optimizes a general model using samples from multiple tasks (a.k.a., meta-tasks) in order to adapt to new unseen tasks. By designing meta-tasks associated with specific communication scenarios using the model-agnostic meta-learning (MAML) framework \cite{finn2017model}, prior work reported better performance for meta-learning solutions compared to standard DL methods. A few attempts bypass the black-box nature of DL models and connect communication systems models with meta-learning. This enables learning a subset of the model-based parameters, thereby minimizing the search space induced by black-box DL models \cite{raviv2023online}. However, the large body of research about generalization in wireless is limited to creating multiple meta-tasks with different communication conditions following the MAML-like framework. While enumerating and generating meta-tasks do not scale for some complex communication scenarios, understanding which feature is invariant across different domains becomes critical for scaling DL techniques. By doing so, one relies on the domain knowledge in addition to the standard two-step optimization of meta-learning. For instance, variations of complex signals related to the phase generalize better than those related to the amplitude. Overall, the study of the features in the context of meta-learning (a.k.a., meta-features) for wireless problems has not been explored by the wireless communications community. Existing studies rely on back-propagation to extract suitable correlations characterizing specific meta-tasks. This is different from  those in the ML community which investigate data properties that affect the learning performance, and measure similarities between datasets and meta-features using multiple criteria such as mutual information and density skewness \cite{rivolli2018characterizing}.

\subsection{Unquestioned Sources of Dataset}

In conventional communications protocols, a significant portion of a transmission interval (e.g., time slot) is usually used to send training sequences for channel estimation. The real-time computation involved during these training time slots defines the computational complexity of many communication methods. For possible future AI-aided communication protocols, it is still unclear whether they will be designed based on fully offline training procedures or by additionally relying on extra finetuning steps. The latter scenario raises the question about real-time collection of data for adaptation purposes, and the related overhead to determine the complexity of DL techniques. The existing research in the open literature disregards these practical challenges and focuses entirely on fully offline-trained DL methods, which, at the current state, cannot fulfill the adaption capabilities of wireless systems envisioned for 6G.

When accuracy is the only metric of evaluation of DL models, ignoring these practical challenges seems acceptable because offline training is enough to judge the DL performance. However, when the assessment of DL techniques accounts for their generalization issues, the availability of data sources and their properties becomes a central component of the analysis.

\subsection{Use of Reinforcement Learning for Optimization}

In the pre-DL era (i.e., before 2012), optimizing non-convex problems using gradient descent (GD) algorithms were not very popular in the wireless research community. Instead of using GD-based optimization, researchers convexified the non-convex problems in order to solve them. DL research has introduced a plethora of optimizers to train DNNs which has made GD  a popular technique. This is partly due to the unique properties of flat minima characterizing DNNs' loss functions \cite{mulayoff2020unique}. Consequently, DL techniques have been used to solve optimization problems such as channel and power allocation, beamforming/precoding, user association, and trajectory planning for unmanned aerial vehicles (UAVs) in the physical/link layers in various contexts and system models.

A common practice to tackle non-convex communication problems in the DL era is by resorting to the reinforcement learning (RL) paradigm. RL agents are trained to learn the optimal policy to act on an environment in order to maximize the sum of the instantaneous reward signals received from an environment. We refer the reader to \cite{sutton2018reinforcement} for a rigorous treatment of RL formulation in terms of Markov decision processes (MDPs). By substituting the reward signal within the RL formalism with a convex or non-convex cost function to be optimized, the RL agent can find the best policy to optimize it. For instance, one can associate the beamforming vector to the action of the RL agent and the achievable rate of the communication system to be the reward signal \cite{al2020multiple}.

Because wireless communications problems must satisfy multiple constraints simultaneously (e.g., power, latency, signal-to-interference-plus-noise ratio [SINR]), prior work made use of clipping strategies to enforce constraints on the RL agent's output. While this strategy provides good  results in some scenarios, it does not guarantee an optimal solution. A better choice would be to cast the communication problem within a constrained MDP formalism \cite{altman1999constrained}. However, little effort has been dedicated to properly incorporating the constraints, and the unconstrained MDP approach remains a popular data-driven approach to optimize non-convex communication problems. In addition, the discussion of RL challenges in terms of sample efficiency and generalization is usually neglected despite the fact that it is an active research area within the ML research community.

\begin{figure*}[!b]
\vspace{-0.1cm}
\centering
\subfloat[]{\includegraphics[scale=0.75]{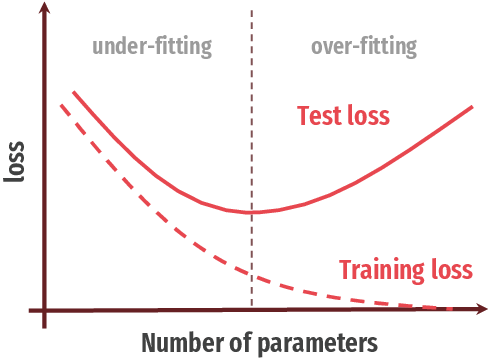}%
\label{fig:classical-vb}}
\hfil
\subfloat[]{\includegraphics[scale=0.75]{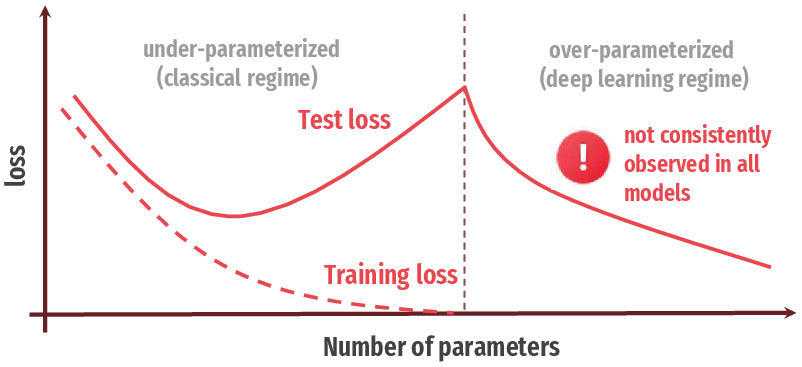}%
\label{fig:deep-vb}}
\caption{Plots of training and test losses characterized by: (a) a U-shaped loss curve justified by the bias-variance trade-off, and (b) the double-descent loss curve combining the U-shaped loss curve for the under-parameterized regime and a decreasing loss for the over-parameterized regime. The latter is not a phenomenon exhibited by all deep learning models and hence is not well understood yet.}
\label{fig:vb-tradeoff}
\end{figure*}
\subsection{Fixed Structures of Deep Neural Networks}

DNNs have fixed structures and it is not possible to change them anymore after initialization. This inflexibility is considered a drawback because many wireless communications problems may require different input/output sizes over time. As one example, consider the problem of channel estimation at the base station based on the SINR vector with each component being associated with a specific user. In this case, the DNN input is the SINR vector while the estimated channel matrix represents the DNN output. During multiple transmission blocks, the number of users changes and so does the size of the SINR vector. Using the dropout method \cite{srivastava2014dropout}, it is possible to randomly drop the contribution of some neurons during the DNN training. While this provides flexibility in the structure of hidden layers, the input and output layers still have a fixed size, and hence limited flexibility in practice. Actively altering the network structure by adding neurons is both a promising and challenging direction. Because DL research for wireless communications generally considers DNNs as black-box modules, this field has not attracted the attention of the wireless research community and opted for constant-padding the input and output layers to the expected maximum size. Another unexplored area in this direction is the mapping of different vector sizes to the same latent space, thereby unifying the DNN input into the same feature space. While feature extractors for computer vision and natural language processing (NLP) tasks are abundant, wireless communications problems require signal-based feature extractors, which have not been well examined by our research community yet. It is worth noting that the recent wave of NLP models has allowed DNNs to have an \textit{apparent} flexibility in the input and output sizes since LLMs are trained to sequentially predict the next token given the current token. This is to be opposed to the application of DL for wireless where the output of DNNs is obtained with a one-shot inference run.

\section{Two Key Trade-offs for DL\\ Applied to Wireless Physical Layer}
In this section, we highlight the importance of two fundamental learning trade-offs in the assessment of DL models, namely, {\em accuracy versus generalization}, and {\em compression versus latency}. In particular, we discuss how accuracy-generalization should guide the evaluation of smaller DL models and highlight the importance of the compression-latency trade-off for practical use cases of LLMs. Through this section, generalization does not refer to the stability of the model's performance under noise and adversarial examples (i.e., adversarial robustness), but rather to the ability of the model to generalize to unseen scenarios.

\begin{figure}[h!]
\centering
\includegraphics[scale=0.6]{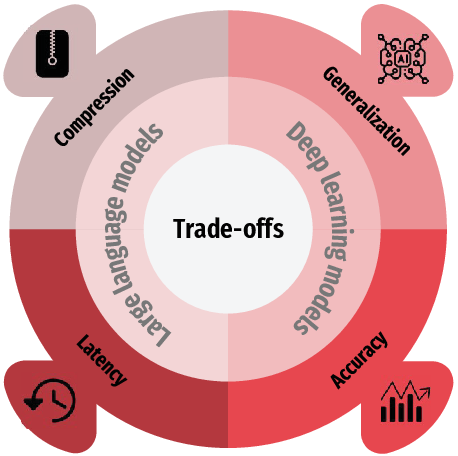}
\caption{The key trade-offs in assessing the relevance of small DL models and LLMs for communication use cases.}
\label{fig:tradeoffs}
\vspace{-0.3cm}
\end{figure}

\subsection{Accuracy Versus Generalization}

\begin{figure*}[!b]
\centering
\subfloat[]{\includegraphics[scale=0.6]{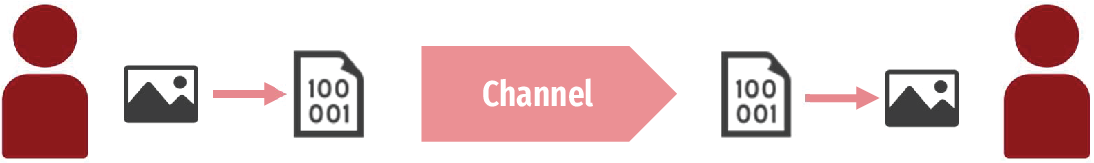}%
\label{fig:classical-communication}}
\hfil
\subfloat[]{\includegraphics[scale=0.6]{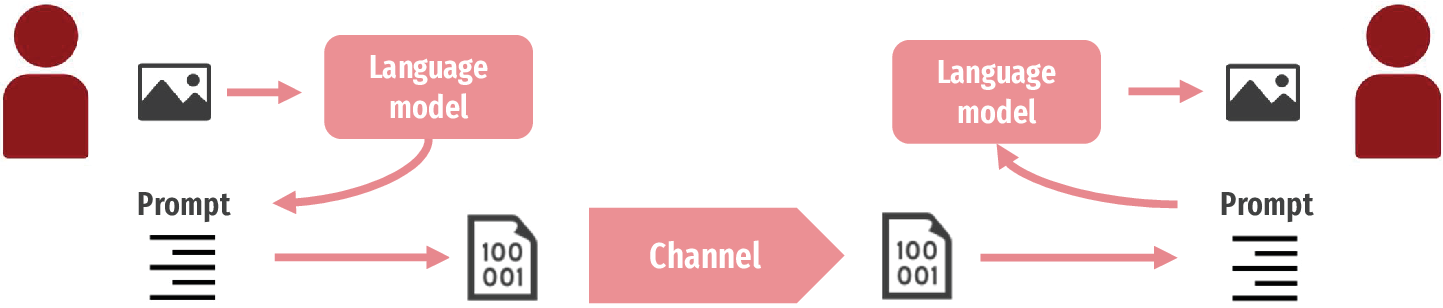}%
\label{fig:gpt-communication}}
\caption{Image transfer using (a) conventional communication, and (b) using a LLM as an image compressor to a prompt.}
\label{fig:classical-gpt-communication}
\end{figure*}

In classical machine learning, the trade-off between the accuracy and the generalization pertains to the fundamental \textit{bias–variance trade-off} which characterizes the generalization capabilities of predictive models \cite{hastie2009elements}. The bias–variance trade-off stipulates that a model must have enough parameters to capture the underlying structure of the dataset without over-fitting spurious patterns. Fig. \ref{fig:classical-vb} depicts the expected U-shaped variation of the test loss whose minimum represents the sweet spot in terms of the number of parameters between under-fitting and over-fitting. In DL practice, however, deeper networks with a large number of parameters are trained to interpolate between the training samples and do maintain a lower test loss accuracy on test datasets as depicted by the ``double descent'' curve in Fig. \ref{fig:deep-vb}. While this over-parameterized regime depicted might look contradictory to classical (i.e., under-parameterized) regime, this over-parameterized regime is not well understood and is being actively investigated within the theoretical DL community \cite{belkin2019reconciling}. This is because the double descent behavior does not consistently occur for every DNN and some of them, even very deep ones, still empirically obey the bias-variance trade-off \cite{bubeck2023universal}. Whether the DNN follows the variance-bias trade-off or the double descent behavior depends on the DNN's parameters such as the number of input data points, the number of layers, and the overall number of parameters.

For the above reasons, evaluating the performance of DL models must be entirely tied to the employed DNN architecture. For example, one cannot claim that DL decoding is superior to classical methods because the mean-square error of \textit{one specific} DNN tested on a few evaluation scenarios is lower than the one of a classical method (cf. Section \ref{subsec:single-metric-eval}). Recent efforts from the communication community initiated the application of different learning paradigms including transfer learning, meta-learning, and continual learning to investigate the generalization of DNNs \cite{akrout2023domain}. Some studies also showed that even the incorporation of domain knowledge in supervised learning without the explicit use of other learning paradigms can improve the DNN's generalization. For instance, by alternating between the time-domain and frequency-domain representations of signals and using the idea of successive-estimation and cancellation, it is possible to design DNNs to handle multiple sinusoid waves and improve their estimation performance on out-of-distribution samples \cite{dreifuerst2022signalnet}.

In summary, by blinding applying DNNs without accounting for their accuracy-generalization trade-off, most research results become biased and less impactful in the long term.

\noindent At the time of writing, potential leaks about the unknown GPT-4 architecture on Reddit \cite{gpt4leak} revealed that GPT-4 is not one giant monolithic lossy dataset compressor but rather an ensemble of eight 220B-parameter LLMs, each trained with different data/task distributions. This suggests that GPT-4 is a mixture of experts (unlike GPT-3.5 and GPT-3) and operates at a much smaller over-parameterized regime that challenges the myth of very large and fully end-to-end DNNs.

\subsection{Compression Versus Latency}\label{subsec:compression-latency}

\begin{figure*}[t!]
\centering
\includegraphics[scale=0.6]{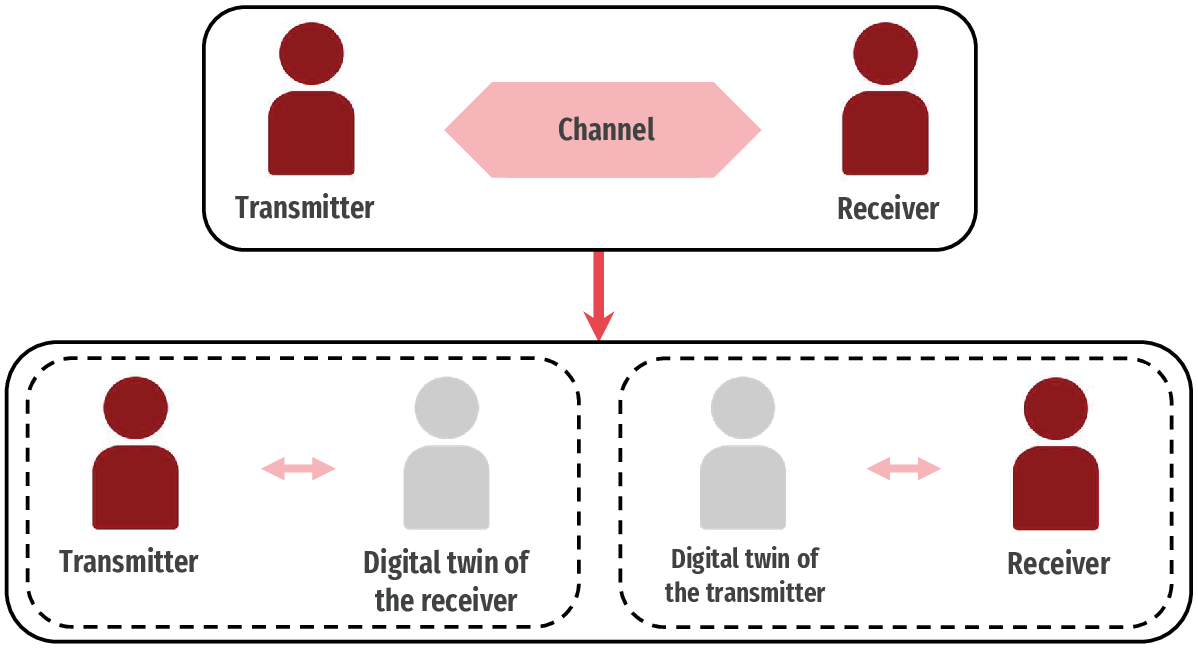}
\caption{The implementation of digital twins depends on finding the trade-off between latency and compression that satisfies the QoS.}
\label{fig:digital-twins}
\vspace{-0.2cm}
\end{figure*}

The study of the trade-off between lossless/lossy compression and latency has historically evolved around reaching the source coding limit established in 1948 by Shannon's seminal work \cite{shannon1948mathematical}. In a nutshell, it stipulates that higher compression ratios require longer block lengths and thus higher encoding and decoding time complexity. Specifically, Shannon showed that the codeword length of an optimal prefix-free code is approximately the negative logarithm of the codeword's probability. He also proved that the expected message length of an optimal prefix-free code is close to the entropy of the message. Shannon also explored the information-theoretic relationship between compression and (next-letter) prediction by estimating the entropy of the English language \cite{shannon1951prediction}. Similar to the connection between next-letter predictors and data compressors, the emerging LLMs represent lossy data compressors in a data-driven fashion which exploit the inherent redundancy within the human language to embed massive datasets into a significantly smaller DNN model \cite{jiang2023low}. They are giant DNNs with billions of parameters compressing a large number of datasets by learning to predict the next token from the previous context composed of one or multiple tokens.

Given the tight information-theoretic relationship between the compression ratio and the decoding latency, it is therefore natural at this time to review some of the envisioned communication use cases of LLMs. The fascination of the general public with the quality of the output text of LLMs suggests that their compression quality will be at the cost of their decoding latency. Indeed, an LLM with a few billion parameters requires around 100 GB of RAM. Moreover, the current optimized deployment of LLMs for efficient inference comes at the expense of high backend infrastructure costs and significant latency. Aware of these issues, important research efforts from the LLM community are actively examining the effect of quantization on the inference time of LLMs without significantly affecting their accuracies. For these reasons, the discussed applications of LLMs by the communications research community focus on latency-tolerant use cases at the application layer.

The first use case of LLMs for communication is image transfer. As shown in Fig. \ref{fig:gpt-communication}, a user equipped with an LLM obtains the text prompt of the image to transfer. Only the prompt is sent through the channel as a bit stream. The prompt is then recovered from the received bits to probe the LLM, which in turn outputs the image to the receiver. This is unlike the conventional data transfer where the entire bit stream of the image is transmitted through the channel as in Fig. \ref{fig:classical-communication}. Here, the LLM-aided image transfer can provide a significant data compression rate under two conditions. The first one is that the parameters of the transmit and receive LLMs must be equal, or that their embeddings yield the encoded and decoded image. 
The second one is related to the compression-latency trade-off where both the encoding and decoding time of LLMs must provide a significant advantage over the improvement of the compression ratio between an image and a text prompt. 

Another use case of LLMs is to minimize the need for communication between two users by learning a personalized LLM as a digital twin for each one of them as shown in Fig. \ref{fig:digital-twins}. This means that an LLM must be able to reliably mimic the user interaction. From a probabilistic perspective, this corresponds to reliably learning the joint distribution of reactions, preferences, and thoughts of any user. The realizability of this scenario is still far from being a reality in the near future due to multiple reasons. This includes the data privacy and security concern, as well as the hallucination effects exhibited by LLMs as they tend to output text that appears to be correct but is actually false or not based on the input given. In summary, short-cutting the cost of communication must come with the benefit of highly accurate digital twins.
\noindent Another important limitation to implementing these use cases is the lack of rigorous metrics to assess the performance of LLMs. In fact, existing evaluation protocols are either automatic (e.g., the ROUGE metric for text summarization) exhibiting poor correlation to human judgments, or manual yielding noisy and potentially biased annotations. In summary, the adaptation of LLMs to the unique signal-based nature of communication datasets is vague and uncertain and is still in its infancy.

\section{Conclusion}
The ever-increasing requirements for future 6G wireless applications call for the urgent need to go beyond the accuracy metric to assess DL methods for communication. We have described the importance of the accuracy-generalization and compression-latency trade-offs in shaping the future evaluation guidelines of DL techniques for wireless problems as summarized in Fig. \ref{fig:tradeoffs}. 
\begin{figure}[h!]
\centering
\includegraphics[scale=0.4]{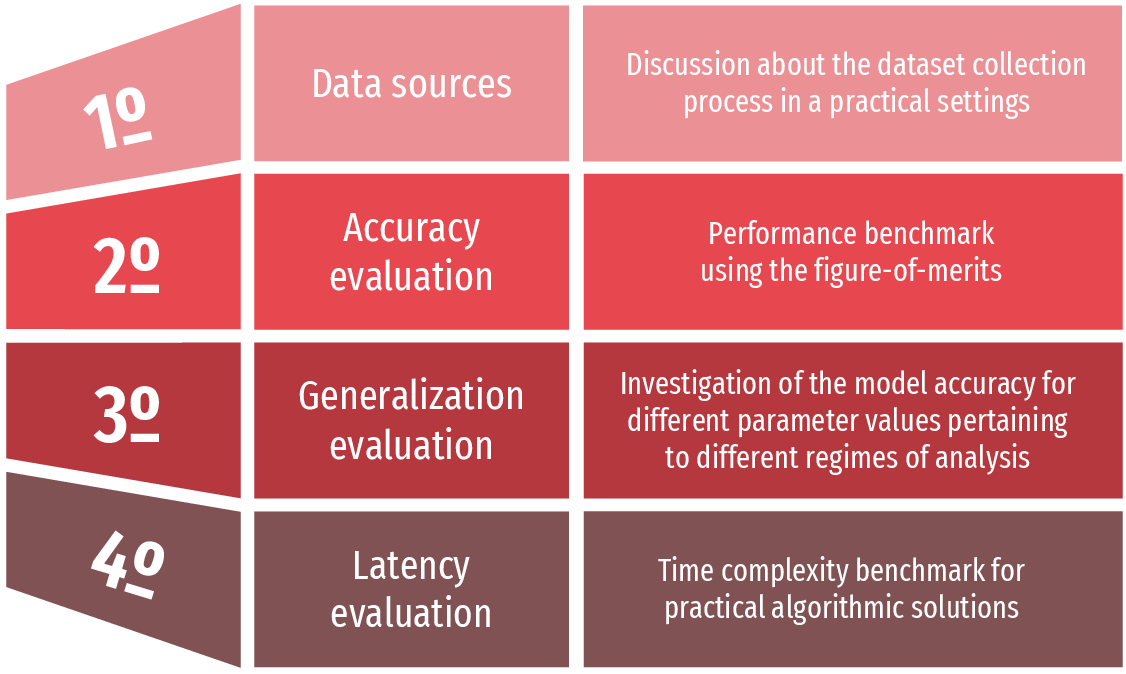}
\caption{The proposed evaluation components.}
\label{fig:tradeoffs}
\end{figure}

\noindent These evaluation criteria should be continuously updated in light of a new understanding of the specific challenges facing the applications of deep learning models for wireless communication problems. We also have discussed how these metrics are critical in evaluating the relevance of emerging deep learning models including large language models. We believe these trade-offs bridge the gap between the empirical nature of deep learning models applied to communication problems and the challenging technical requirements of future communication systems.

\section*{Acknowledgments}
The work of M. Akrout was supported by the doctoral scholarship of the Natural Sciences and Engineering Research Council of Canada (NSERC). The work of A. Mezghani, E. Hossain, and F. Bellili were supported by Discovery Grants from NSERC. The work of R. Heath was supported by the National Science Foundation (grant nos. NSF-ECCS-2153698, NSF-CCF-2225555, NSF-CNS-2147955) and in part by funds from federal agency and industry partners as specified in the Resilient \& Intelligent NextG Systems (RINGS) program. The authors also acknowledge the insightful comments and suggestions from Osvaldo Simeone.

\bibliographystyle{IEEEtran}
\bibliography{IEEEabrv,references}

\end{document}